\documentclass[11pt]{article}
\usepackage{epsfig}
\usepackage{amsmath}
\usepackage{graphicx}
\usepackage{cite}
\usepackage{float}

\begin{document}

\author{S. Dev\thanks{dev5703@yahoo.com} $^{a, b}$, Radha Raman Gautam\thanks{gautamrrg@gmail.com} $^a$ and Lal Singh\thanks{lalsingh96@yahoo.com} $^a$}
\title{Charged Lepton Corrections to Scaling Neutrino Mixing}
\date{$^a$\textit{Department of Physics, Himachal Pradesh University, Shimla 171005, INDIA.}\\
\smallskip
$^b$\textit{Department of Physics, School of Sciences, HNBG Central University, Srinagar, Uttarakhand 246174, INDIA.}}

\maketitle
\begin{abstract}
Assuming Majorana nature of neutrinos, a general expression for the charged lepton corrections to scaling neutrino mixing has been obtained in the context of three flavor neutrino oscillations. Non-zero value of the reactor mixing angle is nicely accommodated. It is noted that scaling in the effective neutrino mass matrix is equivalent to the presence of two vanishing minors corresponding to first row elements of the effective neutrino mass matrix. A value of reactor mixing angle which is fairly close to the currently measured best fit is predicted for charged lepton corrections of the order of the Cabbibo angle. We, also, present symmetry realization of such texture structures in the framework of type-I seesaw mechanism with non-diagonal charged lepton mass matrix using discrete Abelian flavor symmetry.
\end {abstract}

\section{Introduction}
The discovery of neutrino oscillations \cite{experiments} is one of the major discoveries in particle physics which provided the first hint of non-zero neutrino masses and mixing leading to physics beyond the Standard Model (SM). Considerable progress has been made in precise determination of neutrino masses and mixing during the recent past. Recently, a number of neutrino oscillation experiments \cite{t2k, minos, dchooz, dayabay, reno} have established a non-vanishing and relatively large value of reactor neutrino mixing angle ($\theta_{13}$) with a best fit value around $9^{\circ}$. The relatively large value of $\theta_{13}$ has opened the possibilities to explore Dirac-type CP-violation in the lepton sector, distinguishing neutrino mass hierarchy and identifying the octant of atmospheric neutrino mixing angle ($\theta_{23}$). A non-zero value of $\theta_{13}$ has made it necessary to modify mixing schemes like Tribimaximal mixing \cite{tbm} and others resulting from some underlying flavor symmetries which predict $\theta_{13}=0$. One of the ways to achieve a deviation of $\theta_{13}$ from zero is to consider charged lepton corrections and many attempts have been made in the past \cite{chlep1} and, especially, very recently \cite{chlep2} to achieve the deviation of $\theta_{13}$ from zero in terms of charged lepton corrections. In addition, other theoretical ideas such as texture zeroes \cite{tz}, vanishing minors \cite{zerominor,lashin}, hybrid textures \cite{hybrid}, equalities between mass matrix elements or cofactors \cite{tec} have been proposed which, naturally, accommodate a non-zero $\theta_{13}$. \\
One of the attempts to identify the structure of the effective neutrino mass matrix is the Strong Scaling Ans\"{a}tze (SSA) proposed by Mohapatra \textit{et al.} \cite{scaling1}. In SSA \cite{scaling2}, the elements of the effective neutrino mass matrix ($M_\nu$) are related as:  
\begin{eqnarray}
\frac{m_{\mu e}}{m_{\tau e}}=\frac{m_{\mu \mu}}{m_{\tau \mu}}=\frac{m_{\mu \tau}}{m_{\tau \tau}}= s
\end{eqnarray}   
where $s$ is the scaling factor. The scaling neutrino mass matrix has the following form
\begin{equation}
 M_\nu = \left(
\begin{array}{ccc}
a & b & b/s \\ b & d & d/s \\ b/s & d/s & d/s^2
\end{array}
\right)
\end{equation}
which is singular and has a vanishing eigenvalue. The eigenvector associated with the vanishing eigenvalue is 
$(0,-\frac{1}{\sqrt{1+s^2}},\frac{s}{\sqrt{1+s^2}})^T$. Hence, the above mass matrix is only compatible with inverted neutrino mass hierarchy along with a vanishing mass eigenvalue and predicts a vanishing reactor mixing angle implying the absence of Dirac-type CP-violation in the lepton sector.
However, in the light of recent experimental results on $\theta_{13}$ one needs suitable modifications to the SSA to explain a non-zero value of $\theta_{13}$. To accomplish this, deviations from SSA have been considered recently by adding some scaling breaking parameters to the effective neutrino mass matrix elements \cite{scaling3}.\\
Equation (1) implies the following relations between the neutrino mass matrix elements
\begin{align}
m_{\mu \mu} m_{\tau \tau} - m_{\tau \mu} m_{\mu \tau}&=0,  \nonumber \\
m_{\mu e} m_{\tau \tau} - m_{\tau e} m_{\mu \tau}&=0, \\ \nonumber
\textrm{and}\ \ \ \ \ m_{\mu e} m_{\tau \mu }- m_{\tau e} m_{\mu \mu} & =0
\end{align} 
which correspond to vanishing minors of the first row elements of the neutrino mass matrix. These three relations are not independent and it can be easily seen that the vanishing of any set of two minors leads to a vanishing third minor. The interesting fact that has been completely overlooked in the literature is that scaling in the effective neutrino mass matrix is equivalent to two vanishing minors corresponding to the first row of the effective neutrino mass matrix.  In the flavor basis, all possible cases of two vanishing minors in the neutrino mass matrix have been studied in \cite{lashin}. The cases $S_1, S_2, S_3$ in the nomenclature of \cite{lashin} correspond to the SSA. In recent works, where deviations from the SSA are considered to accommodate non-zero value of $\theta_{13}$ by introducing some scaling breaking parameter in the neutrino mass matrix elements, other predictions of scaling are also lost. For example, the mass eigenvalue $m_{3}$ becomes non-vanishing and one has to do away with the prediction that only inverted hierarchy of neutrino masses is satisfied. \\
In the present work, we study charged lepton corrections to the scaling neutrino mixing in terms of two vanishing minors in the first row of the neutrino mass matrix, thereby, accommodating the non-zero value of $\theta_{13}$ with the advantage that such corrections will only affect the neutrino mixing and CP-violation whereas some scaling properties like the inverted hierarchy and vanishing $m_{3}$ remain intact. One of the most elegant mechanisms to understand the smallness of neutrino masses is the type-I seesaw mechanism \cite{seesaw1} in which one extends the SM by adding heavy right-handed neutrinos. In the framework of type-I seesaw mechanism, the effective Majorana neutrino mass matrix is given by 
\begin{equation}
M_\nu \approx -M_D M_R^{-1} M_D^T
\end{equation}
where $M_D$ is the Dirac neutrino mass matrix and $M_R$ is the right-handed Majorana neutrino mass matrix. In the context of type-I seesaw mechanism, scaling in the effective neutrino mass matrix can be obtained by considering a diagonal $M_R$ with $M_D$ having the form
\begin{equation}
 M_D = \left(
\begin{array}{ccc}
A & 0 & B \\ 0 & 0 & C \\ 0& 0 & D
\end{array}
\right)
\end{equation}
which, by no means, is the only possibility. We present the symmetry realization of above $M_D$ and $M_R$ in the non-diagonal charged lepton basis using discrete Abelian flavor symmetry $Z_{6}$.

\section{Formalism}
A complex symmetric Majorana neutrino mass matrix is diagonalized by a unitary matrix $U_{\nu}$ as
\begin{equation}
M_{\nu}= U_{\nu} M_{\nu}^{diag}U_{\nu}^{T}
\end{equation}
where $M_{\nu}^{diag}$ = diag$(m_1,m_2,m_3)$. The lepton mixing matrix ($U_{PMNS}$) can be written in terms of the product of two 3$\times$3 unitary matrices $U_l$ and $U_\nu$ arising from diagonalization of the charged lepton mass matrix ($M_l$) and the neutrino mass matrix ($M_\nu$), respectively \cite{pmns}:
\begin{equation}
U_{PMNS}= U_{l}^{\dagger} U_{\nu}
\end{equation} which can be, further, parametrized as
\begin{equation}
U_{PMNS} = \mathcal{P} V \widehat{P}
\end{equation}
where  \cite{foglipdg}
\begin{equation}
V= \left(
\begin{array}{ccc}
c_{12}c_{13} & s_{12}c_{13} & s_{13}e^{-i\delta} \\
-s_{12}c_{23}-c_{12}s_{23}s_{13}e^{i\delta} &
c_{12}c_{23}-s_{12}s_{23}s_{13}e^{i\delta} & s_{23}c_{13} \\
s_{12}s_{23}-c_{12}c_{23}s_{13}e^{i\delta} &
-c_{12}s_{23}-s_{12}c_{23}s_{13}e^{i\delta} & c_{23}c_{13}
\end{array}
\right)
\end{equation} with $s_{ij}=\sin\theta_{ij}$ and $c_{ij}=\cos\theta_{ij}$, $\mathcal{P}=$ diag$(e^{i\phi_{1}},e^{i\phi_{2}},e^{i\phi_{3}})$ and
$\widehat{P} =$ diag$(1,e^{i\alpha},e^{i(\beta+\delta)})$ where the diagonal phase matrix $\mathcal{P}$ is physically unobservable. However, one needs to take into account the phase matrix $\mathcal{P}$ to avoid parameter mismatch \cite{xing}. $\widehat{P}$ is the diagonal phase matrix with one Dirac-type CP-violating phase $\delta$ and two Majorana-type CP-violating phases $\alpha$, $\beta$. A general unitary 3$\times$3 matrix can be parametrized in terms of three mixing angles and six phases as \cite{pascoli}
\begin{equation}
U=P \tilde{U} Q,
\end{equation}
where $P=$ diag$(e^{i\gamma_{1}},e^{i\gamma_{2}},e^{i\gamma_{3}})$ and $Q=$ diag$(1,e^{i\rho_{1}},e^{i\rho_{2}})$ are diagonal phase matrices and $\tilde{U}$ has the form similar to $V$ as given in equation (9). $M_l$ can be diagonalized by a bi-unitary transformation: 
\begin{equation}
M_{l}=U_{l}M^{diag}_{l}U^{\dagger}_{R}. 
\end{equation}
The Hermitian product $M_{l}M^{\dagger}_{l}$, therefore, becomes
\begin{equation}
M_{l}M^{\dagger}_{l}=U_{l}M^{diag}_{l}U^{\dagger}_{R}U_{R}M^{diag}_{l}U^{\dagger}_{l}
=U_{l}(M^{diag}_{l})^{2}U^{\dagger}_{l}
\end{equation}
where $U_l=P^{l} \tilde{U}^{l}$. For individual $i$-$j$ sectors $(i<j=1,2,3)$, $\tilde{U}^{l}$ becomes a real orthogonal rotation matrix corresponding to rotation in that particular sector.
Using equations (6) and (7), the effective neutrino mass matrix can be written as \cite{fourzero}
\begin{equation}
M_{\nu}=U_l U_{PMNS} M_{\nu}^{diag}U_{PMNS}^{T} U_l^T.
\end{equation}
The Dirac-type CP-violation in neutrino oscillation experiments can be described through a rephasing invariant quantity $J_{CP}$ \cite{jarlskog} which, in the parametrization adopted here, is given by
 \begin{equation}
J_{CP} = s_{12}s_{23}s_{13}c_{12}c_{23}c_{13}^2 \sin \delta \   .
\end{equation}   
\section{Charged Lepton Corrections and Scaling Neutrino Mass Matrix}
In this section, we first study the effects of charged lepton corrections coming from $i$-$j~(i<j=1,2,3)$ sectors individually along with scaling in the neutrino mass matrix. The analysis is done by considering two vanishing minors corresponding to the elements of the first row of the  neutrino mass matrix and then substituting $m_{3}=0$. We, then, obtain the most general results for the combined contribution of charged lepton sector to the reactor neutrino  mixing angle $\theta_{13}$ and Dirac-type CP-violating phase $\delta$.
\subsection{Contribution from 2-3 sector}
For contributions coming from individual charged lepton sectors, equation (13) can be written as
\begin{align}
M_{\nu}&=P^{l} O^{l}_{ij} U_{PMNS} M_{\nu}^{diag}U_{PMNS}^{T} {O^{l}_{ij}}^T{P^{l}}^T
\end{align}
where $O^{l}_{ij}$ represents the real orthogonal mixing matrix of the $i$-$j$ sector
\begin{small}
\begin{equation}
O_{12}^l = \left(
\begin{array}{ccc}
\cos{\theta_{12}^l} & \sin{\theta_{12}^l} & 0 \\ -\sin{\theta_{12}^l} & \cos{\theta_{12}^l}& 0 \\ 0 & 0 & 1
\end{array}
\right) , \ \ \ 
O_{13}^l = \left(
\begin{array}{ccc}
\cos{\theta_{13}^l} & 0 & \sin{\theta_{13}^l} \\ 0 & 1 & 0 \\ -\sin{\theta_{13}^l} & 0 & \cos{\theta_{13}^l}
\end{array}
\right), \ \ \ 
O_{23}^l = \left(
\begin{array}{ccc}
1 & 0 & 0 \\ 0 & \cos{\theta_{23}^l} & \sin{\theta_{23}^l} \\ 0 & -\sin{\theta_{23}^l} & \cos{\theta_{23}^l}
\end{array}
\right).
\end{equation}
\end{small}
For contribution from the 2-3 sector of charged leptons, equation (15) has the following form:
\begin{equation}
M_{\nu}=P^{l} O^{l}_{23} U_{PMNS} M_{\nu}^{diag}U_{PMNS}^{T} {O^{l}_{23}}^{T}{P^{l}}^T.
\end{equation}
Simultaneous existence of two vanishing minors corresponding to the first row of the effective neutrino mass matrix with $m_3 = 0$ implies a common vanishing factor
 \begin{equation}
 s_{13} e^{i (\delta +\phi_{2}+\phi_{3})}=0
 \end{equation}
which is independent of the contribution from the 2-3 sector of charged leptons. The reactor mixing angle remains zero and there is no Dirac-type CP-violation in this case as $J_{CP}=0$. However, the atmospheric mixing angle $\theta_{23}$ gets affected because of the change in the scaling factor due to contribution from charged lepton sector \cite{scaling1}.
\subsection{Contribution from 1-2 sector}
When only the 1-2 sector of charged leptons contributes to lepton mixing, the effective neutrino mass matrix can be written as
\begin{equation}
M_{\nu}=P^{l} O^{l}_{12} U_{PMNS} M_{\nu}^{diag}U_{PMNS}^{T} {O^{l}_{12}}^{T}{P^{l}}^T.
\end{equation}
Following the procedure outlined above, the condition of two vanishing minors corresponding to the first row of $M_\nu$ reduces to
\begin{align}
c_{12}^{l} s_{13} \cos {(\delta +\phi_{2}+\phi_{3})} + c_{13} s_{12}^{l} s_{23} \cos{(\phi_{1}+\phi_{3})}&=0,\\
c_{12}^{l} s_{13} \sin{(\delta +\phi_{2}+\phi_{3})} + c_{13} s_{12}^{l} s_{23} \sin{(\phi_{1}+\phi_{3})}&=0,
\end{align}
which implies
\begin{align}
&\tan{\delta}=\tan{(\phi_{1}-\phi_{2})},  \\
&\sin{\theta_{13}}=\frac{\sin{\theta_{23}} \ s_{12}^{l}}{\sqrt{1-(s_{12}^{l} \cos{\theta_{23}})^{2}}} .
\end{align}
From equation (22), the phase difference $(\phi_{1}-\phi_{2})$ can be identified as the Dirac-type CP-violating phase in the PDG representation \cite{foglipdg} and equation (23) implies a non-zero $\theta_{13}$  resulting from the contribution from the 1-2 charged lepton sector. The correlation plot between $\theta_{13}$ and $\theta_{23}$ is given in Fig.1(a). The present experimental constraints \cite{valledata} limit charged lepton contribution $s_{12}^{l}$ to the range (0.15 - 0.29). The variation of $\theta_{13}$ with $s_{12}^l$ has been depicted in Fig.1(b). For a maximal atmospheric neutrino mixing angle i.e. $s_{23}=\frac{1}{\sqrt{2}}$ and charged lepton contribution $s_{12}^{l}=0.223$ which is of the order of Cabbibo angle, we get a reactor neutrino mixing angle $\theta_{13}=9.19^{\circ}$ close to the current experimental best fit value. The current best fit value of $s_{23}$ with $s_{12}^{l}=0.223$ predicts a reactor mixing angle $\theta_{13} = 10.05^{\circ}$ which is within the 3$\sigma$ range of the current experimental data. The Jarlskog measure of Dirac-type CP-violation `$J_{CP}$' is given by
\begin{equation}
J_{CP}= -\frac{1}{4} \frac{\sin{2\theta_{12}} \sin{2\theta_{23}}~\sin{\theta_{23}}}{ (1-(s^{l}_{12} \cos{\theta}_{23})^{2})^{3/2}}~s^{l}_{12} (c^{l}_{12})^{2}~\sin{(\phi_{1}-\phi_{2})}.
\end{equation}
\begin{figure}[H]
\begin{center}
{\epsfig{file=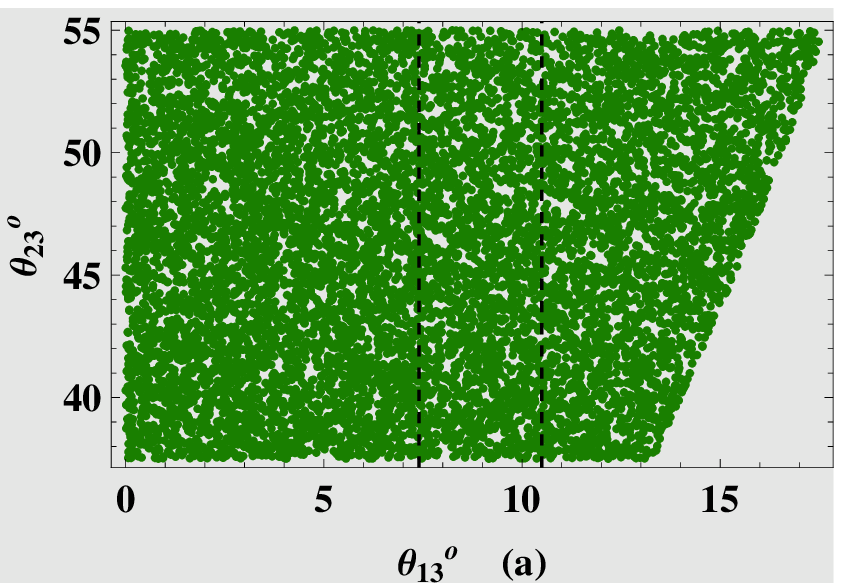, width=5.0cm, height=4.0cm}  \epsfig{file=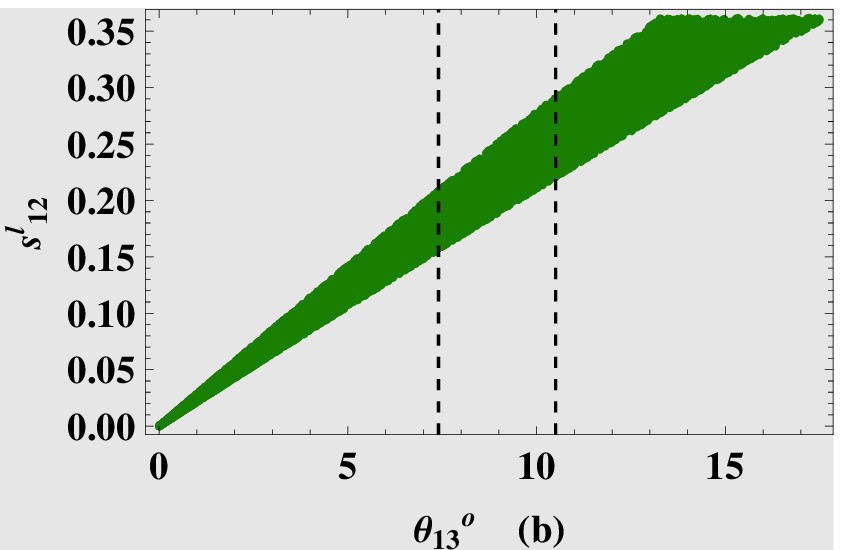, width=5.0cm, height=4.0cm}}
\caption{Correlation plots for charged lepton correction from 1-2 sector. The region between dashed lines is the presently allowed 3$\sigma$ range of $\theta_{13}$.}
\end{center}
\end{figure}
\subsection{Contribution from 1-3 sector}
The effective neutrino mass matrix for charged lepton corrections coming from the 1-3 sector can be written in the form
\begin{equation}
M_{\nu}=P^{l} O^{l}_{13} U_{PMNS} M_{\nu}^{diag}U_{PMNS}^{T} {O^{l}_{13}}^{T}{P^{l}}^T.
\end{equation}
The condition of two vanishing minors corresponding to the first row of $M_\nu$ implies
\begin{align}
c_{13}^{l} s_{13} \cos{(\delta +\phi_{2}+\phi_{3})} + c_{13} c_{23} s^{l}_{13} \cos{(\phi_{1}+\phi_{2})}&=0,\\
c_{13}^{l} s_{13} \sin{(\delta +\phi_{2}+\phi_{3})} + c_{13} c_{23} s_{13}^{l} \sin{(\phi_{1}+\phi_{2})}&=0,
\end{align}
which gives
\begin{align}
&\tan{\delta}=\tan{(\phi_{1}-\phi_{3})}, \\
&\sin{\theta_{13}}=\frac{\cos{\theta_{23}} \ s_{13}^{l}}{\sqrt{1-( s_{13}^{l} \sin{\theta_{23}})^{2}}} .
\end{align}
 From equation (28), one can see that the phase difference $(\phi_{1}-\phi_{3})$ is the same as the Dirac-type CP-violating phase in the PDG representation. The correlation plot between $\theta_{13}$ and $\theta_{23}$ for charged lepton contribution coming from 1-3 sector is given in Fig.2(a). The present experimental constraints limit charged lepton contribution $s_{13}^{l}$ to the range (0.15 - 0.326). The variation of $\theta_{13}$ with $s_{13}^l$ has been depicted in Fig.2(b). For $s_{23}=\frac{1}{\sqrt{2}}$ and a charged lepton contribution $s_{13}^{l}=0.223$ which is of the order of Cabbibo angle, we get a reactor neutrino mixing angle $\theta_{13} = 9.19^{\circ}$ fairly close to the current experimental best fit value. The current best fit value of $s_{23}$ with $s_{13}^{l} = 0.223$ predicts $\theta_{13} = 8.24^{\circ}$ which is within the 2$\sigma$ range of the current experimental data. The Jarlskog measure of Dirac-type CP-violation `$J_{CP}$' is given by
\begin{equation}
J_{CP}= -\frac{1}{4} \frac{\sin{2\theta_{12}} \sin{2\theta_{23}}~\cos{\theta_{23}}}{ (1-(s^{l}_{13} \sin{\theta}_{23})^{2})^{3/2}}~s^{l}_{13} (c^{l}_{13})^{2}~\sin{(\phi_{1}-\phi_{3})}.
\end{equation}
\begin{figure}[H]
\begin{center}
{\epsfig{file=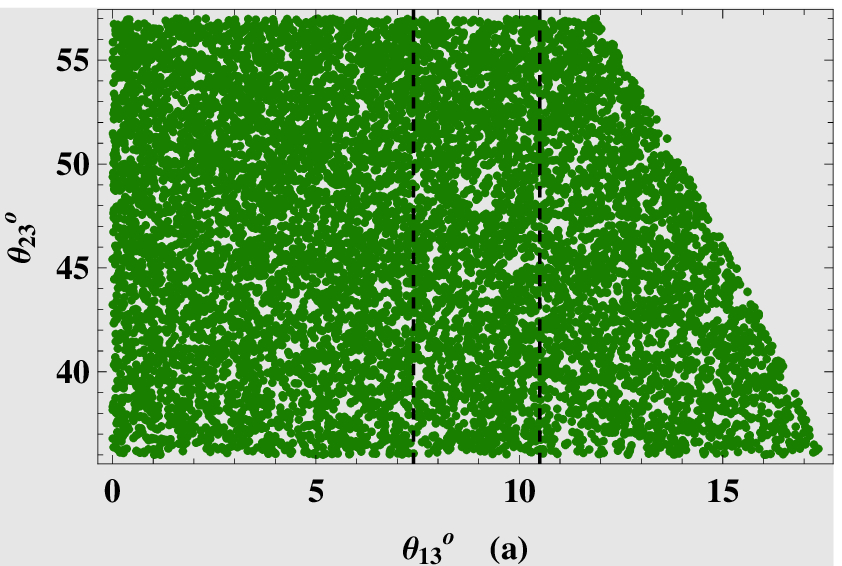, width=5.0cm, height=4.0cm} 
\epsfig{file=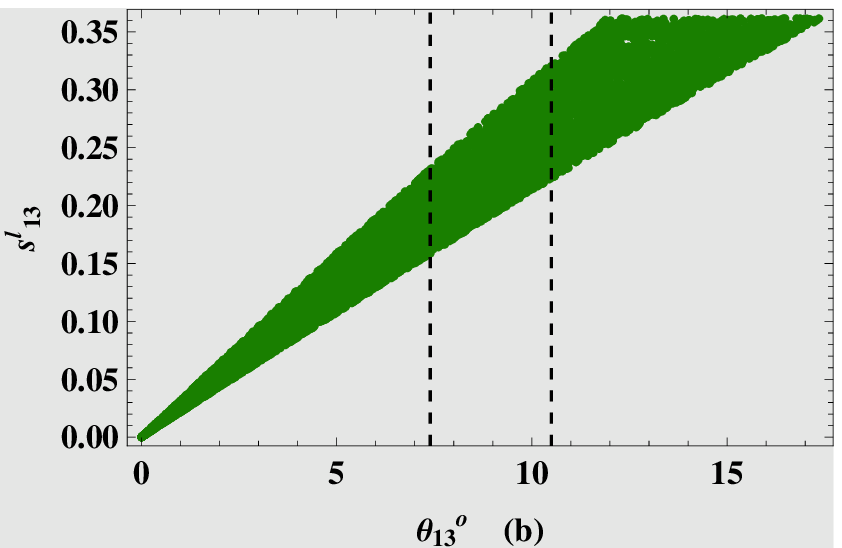, width=5.0cm, height=4.0cm}}
\caption{Correlation plots for charged lepton correction from 1-3 sector. The region between dashed lines is the presently allowed 3$\sigma$ range of $\theta_{13}$.}
\end{center}
\end{figure}
\subsection{General results}
Now we discuss the general case when all $i$-$j~(i<j=1,2,3)$ sectors contribute to lepton mixing.   
Presence of scaling or the simultaneous existence of two vanishing minors corresponding to the first row of the effective neutrino mass matrix with $m_3 = 0$ implies a common vanishing factor:
\begin{align}
c_{13} e^{i \phi_{1}} (c_{23} (\tilde{U}_{(22)}^{l} \tilde{U}_{(31)}^{l}-\tilde{U}_{(21)}^{l} \tilde{U}_{(32)}^{l})&+ s_{23} e^{i \phi_{3}} (\tilde{U}_{(21)}^{l} \tilde{U}_{(33)}^{l}-\tilde{U}_{(23)}^{l} \tilde{U}_{(31)}^{l}))\nonumber \\ 
&+ e^{i (\delta + \phi_{2}+\phi_{3})} (\tilde{U}_{(23)}^{l} \tilde{U}_{(32)}^{l}-\tilde{U}_{(22)}^{l} \tilde{U}_{(33)}^{l}) s_{13}=0,
\end{align}
where $\tilde{U}_{(ij)}^{l}~(i,j=1,2,3)$ denote the elements of $\tilde{U}_{l}$ which on simplification yields
\begin{align}
c_{12}^{l} c_{13}^{l} s_{13} \cos{(\delta+\phi_{2}+\phi_{3})} + c_{13} c_{13}^{l} s_{12}^{l} s_{23} \cos{(\phi_{1}+\phi_{3})}+ c_{13} c_{23} s_{13}^{l} \cos{(\delta^{l}+\phi_{1}+\phi_{2})}&=0,\\
c_{12}^{l} c_{13}^{l} s_{13} \sin{(\delta+\phi_{2}+\phi_{3})} + c_{13} c_{13}^{l} s_{12}^{l} s_{23} \sin{(\phi_{1}+\phi_{3})}+ c_{13} c_{23} s_{13}^{l} \sin{(\delta^{l}+\phi_{1}+\phi_{2})}&=0.
\end{align}
Equations (32) and (33) can, also, be re-expressed as
\begin{align}
a s_{13} \cos{(\delta+\phi_{2}+\phi_{3})} + (b \cos{(\phi_{1}+\phi_{3})} +c \cos{(\delta^{l}+\phi_{1}+\phi_{2})})c_{13}&=0, \\
a s_{13} \sin{(\delta+\phi_{2}+\phi_{3})} + (b \sin{(\phi_{1}+\phi_{3})} +c \sin{(\delta^{l}+\phi_{1}+\phi_{2})}) c_{13}&=0, 
\end{align}
where
\begin{align}
a=c_{12}^{l} c_{13}^{l},~ b= s_{23} c_{13}^{l} s_{12}^{l}~\textrm{and}~ c= c_{23} s_{13}^{l}
\end{align}
are functions of charged lepton mixing angles and atmospheric neutrino mixing angle. Solving equations (34) and (35) leads to the following expressions for $\tan\delta$ and $\sin^{2}{\theta_{13}}$:
\begin{align}
\tan\delta &= \frac{x-\tan(\phi2+\phi3)}{1+x \tan(\phi2+\phi3)}, \\
\sin^{2}{\theta_{13}}&=\frac{b^2+c^2+2bc\cos{(\delta^{l}+\phi_{2}-\phi_{3})}}{{a^2+b^2+c^2+2 bc\cos{(\delta^{l}+\phi_{2}-\phi_{3})}}}
\end{align}
where
\begin{equation}
x = \frac{b \sin{(\phi_{1}+\phi_{3})}+ c \sin{(\delta^{l}+\phi_{1}+\phi_{2})}} {b \cos{(\phi_{1}+\phi_{3})} + c \cos{(\delta^{l}+\phi_{1}+\phi_{2})}}.
\end{equation}
One can, easily, see that for the most general charged lepton corrections, $\theta_{13}$ is independent of the charged lepton contribution from the 2-3 sector (i.e. $\theta_{23}^l$) which is, also, evident from  equation (18). The Jarlskog measure of Dirac-type CP-violation `$J_{CP}$' is given by
\begin{equation}
J_{CP}= -\frac{1}{4} \frac{\sin{2\theta_{12}} \sin{2\theta_{23}}~(b \sin{(\phi_{1}-\phi_{2})} + c \sin{(\delta^{l}+\phi_{1}-\phi_{3})})~a^{2}} {(a^2+b^2+c^2+2 b c \cos{(\delta^{l}+\phi_{2}-\phi_{3})})^{3/2}}.
\end{equation}
For $s_{13}^l = s_{23}^l =0$, equations (37), (38) and (40) reduce to the contributions coming from the 1-2 charged lepton  sector whereas for $s_{12}^l = s_{23}^l =0$ equations (37), (38) and (40) reduce to the contributions coming from 1-3 charged lepton  sector which have been discussed individually in this work.\\
The observation of a non-zero value of the effective Majorana mass of electron neutrino ($M_{ee}$) in neutrinoless double beta (NDB) decay  experiments will establish lepton number violation and the Majorana nature of neutrinos. The effective Majorana mass $M_{ee}$, which determines the rate of NDB decay, is given by
\begin{equation}
M_{ee}=|m_{1}c^{2}_{12}c^{2}_{13}+ m_{2} s^{2}_{12}c^{2}_{13}e^{2i\alpha}+ m_{3} s^{2}_{13}e^{2i\beta}|.
\end{equation}
A large number of projects \cite{cuoricinoexo} aim to achieve a sensitivity upto 0.01 eV for $M_{ee}$. Since in scaling $m_{3}=0$, so only one Majorana phase is present and the expression for $M_{ee}$ in terms of mass squared differences ($\Delta m_{ij}^2 \equiv m_i^2 - m_j^2$) can be written as
\begin{equation}
M_{ee}=\mid c^{2}_{12}c^{2}_{13} \sqrt{\mid \Delta m_{31}^{2}\mid}+s^{2}_{12}c^{2}_{13}e^{2i\alpha} \sqrt{\Delta m_{21}^{2}-\Delta m_{31}^{2}}  \mid .
\end{equation}
The range of $M_{ee}$ for scaling neutrino mass matrix with charged lepton corrections, using the present experimental neutrino oscillation parameters \cite{valledata} is (0.011 - 0.052)eV which is well within the reach of forthcoming NDB decay experiments.
\section{Symmetry Realization}
The symmetry realization of SSA has been studied in the past by many authors \cite{scaling1,scaling2} using discrete non-Abelian flavor symmetries. Since scaling is equivalent to two vanishing minors in the first row of the neutrino mass matrix, it is possible to obtain SSA using discrete Abelian symmetries as two vanishing minors in $M_\nu$ can be realised using discrete Abelian symmetries in the context of type-I seesaw mechanism. In the flavor basis, where the charged lepton mass matrix is diagonal, similar texture structures have, also, been realized in Ref. \cite{lashin} using discrete Abelian flavor symmetries.\\
We, now, present a type-I seesaw realization of the scaling neutrino mass matrix using the discrete Abelian flavor symmetry $Z_{6}$ in the non-diagonal charged lepton basis. To obtain scaling in the effective neutrino mass matrix, one of the simplest possibilities which by no means excludes others is to have the following structures for $M_{D}$ and $M_{R}$ :
 \begin{equation}
 M_D = \left(
\begin{array}{ccc}
A & 0 & B \\ 0 & 0 & C \\ 0& 0 & D
\end{array}
\right),~~
 M_R = \left(
\begin{array}{ccc}
X & 0 & 0 \\ 0 & Y & 0 \\ 0 & 0 & Z
\end{array}
\right)
\end{equation}
which lead to scaling neutrino mass matrix with scaling factor equal to $C/D$. Here, $M_{D}$ has zero textures in a basis where $M_R$ is diagonal. In general, the underlying symmetries for obtaining zero textures in a mass matrix can be the Abelian discrete symmetries \cite{grimus_ab}.\\
In addition to the SM left-handed $SU(2)$ lepton doublets $D_{l L}$ $(l = e, \mu, \tau)$ and the right-handed charged-lepton $SU(2)$ singlets $l_R$, we introduce three right-handed neutrinos $\nu_{l R}$. In the scalar sector, we need three Higgs doublets $\Phi_i$ ($i=1,2,3$) and a scalar singlet $\chi$.
In case of non-diagonal charged lepton contributions from 1-2 sector, we consider the following transformation properties of various fields under $Z_{6}$:
\begin{align}
&\overline{D}_{eL} \rightarrow \overline{D}_{eL},~~~~e_{R} \rightarrow \omega e_{R},~~~~\nu_{eR} \rightarrow \omega^{2} \nu_{eR},\nonumber \\ 
&\overline{D}_{\mu L} \rightarrow \omega \overline{D}_{\mu L},~~\mu_{R} \rightarrow \omega \mu_{R},~~~\nu_{\mu R} \rightarrow \omega^{3} \nu_{\mu R},\nonumber \\
&\overline{D}_{\tau L} \rightarrow \omega^{2}\overline{D}_{\tau L},~\tau_{R} \rightarrow \omega^{4} \tau_{R},~~~\nu_{\tau R} \rightarrow \nu_{\tau R},
\end{align} where  
\begin{eqnarray}
D_{lL} = \left(
\begin{array}{c}
\nu_{lL}   \\ l_{L}  
\end{array}
\right)
\end{eqnarray} 
with $\omega = e^{2 \pi i/6}$ as the generator of $Z_{6}$. 
The bilinears $\overline{D}_{lL}l_{R}$, $\overline{D}_{lL}\nu_{lR}$ and $\nu_{lR}\nu_{lR}$ relevant for $M_{l},~M_{D}$ and $M_{R}$, respectively, transform as
\begin{align}
&\overline{D}_{lL}l_{R} \sim 
\left( \begin{array}{ccc}
w & w & w^{4} \\ w^{2} & w^{2} & w^{5} \\ w^{3} & w^{3} & 1
\end{array}
\right), 
&\overline{D}_{lL}\nu_{lR} \sim 
\left( \begin{array}{ccc}
w^{2} & w^{3} & 1 \\ w^{3} & w^{4} & w \\ w^{4} & w^{5} & w^{2}
\end{array}
\right)
\ \textrm{and}~ \  \nu_{lR}\nu_{lR} \sim 
\left( \begin{array}{ccc}
w^{4} & w^{5} & w^{2} \\ w^{5} & 1 & w^{3} \\ w^{2} & w^{3} & 1
\end{array}
\right).
\end{align}
For contribution to neutrino mixing coming from the 1-2 charged lepton sector, $M_{l}$ and $M_{D}$ require presence of three SU(2) scalar Higgs doublets which transform under $Z_{6}$ as
\begin{align}
\Phi_{1}\rightarrow w^{4}~\Phi_{1},~~~ \Phi_{2}\rightarrow w^{5}~\Phi_{2},~~~ \Phi_{3}\rightarrow ~\Phi_{3}.   
\end{align} 
These transformations give the desired singular Dirac neutrino mass matrix $M_{D}$ of the form given in equation (43) with the following structure of $M_{l}$:
\begin{equation}
M_{l}=\left(
\begin{array}{ccc}
\times & \times & 0 \\
\times & \times & 0 \\
0 & 0  & \times
\end{array}
\right)
\end{equation}
where ``$\times$" represents non-vanishing mass matrix elements. To obtain a diagonal $M_R$, one requires a scalar singlet $\chi$ transforming as $\chi \rightarrow w^{2} \chi$ which couples with the (1,1) element of $M_{R}$ whereas (2,2) and (3,3) elements are the bare mass terms.\\
For charged lepton contribution coming from 1-3 sector only, we consider the following transformation properties of various fields under $Z_{6}$:
\begin{align}
&\overline{D}_{eL} \rightarrow \overline{D}_{eL},~~\overline{D}_{\mu L} \rightarrow \omega^{2} \overline{D}_{\mu L},~~\overline{D}_{\tau L} \rightarrow \omega \overline{D}_{\tau L},\nonumber \\ 
&e_{R} \rightarrow \omega e_{R},~~~~\mu_{R} \rightarrow \omega^{4} \mu_{R},~~~~~\tau_{R} \rightarrow \omega \tau_{R},
\end{align}
 while keeping the transformations of right handed neutrinos ($\nu_{lR}$), the Higgs doublets ($\Phi_i$) and the scalar singlet ($\chi$), the same as in the earlier case. These $Z_6$ assignments lead to the desired structures of $M_{D}$ and $M_{R}$ given in equation (43) with $M_{l}$ having the following structure:
 \begin{equation}
M_{l}=\left(
\begin{array}{ccc}
\times & 0 & \times  \\
0  & \times  & 0 \\
\times  & 0  & \times 
\end{array}
\right).
\end{equation}
Similarly, the symmetry realization for the most general contribution from charged lepton sector can be achieved with an extended Higgs sector and some larger discrete Abelian flavor symmetry.
 \section{Summary}
 To summarize, it is noted that scaling in the effective neutrino mass matrix is equivalent to the presence of two vanishing minors corresponding to the elements of the first row. We utilize this fact to explore the contributions from different sectors of charged leptons to the scaling neutrino mixing and find that the charged lepton contributions from 1-2 and 1-3 sectors result in a non-zero reactor mixing angle and Dirac-type CP-violation. For charged lepton contributions of the order of Cabibbo angle from either 1-2 or 1-3 sectors, the reactor mixing angle is predicted to be fairly close to the current best fit value. The advantage of our approach is that such corrections will only affect neutrino mixing and CP-violation whereas generic predictions of scaling such as inverted hierarchy with a vanishing mass eigenvalue remain intact. The most general results for the reactor mixing angle and the Dirac-type CP-violating phase have been obtained in terms of charges lepton contributions to the scaling neutrino mixing. We, also, present symmetry realization of these texture structures in the context of type-I seesaw mechanism using a discrete Abelian symmetry $Z_6$ where the charged lepton mass matrix is non-diagonal with contributions from either 1-2 or 1-3 sectors and scaling in the neutrino mass matrix.

\textbf{\textit{\Large{Acknowledgements}}}\\
R. R. G.  and L. S. gratefully acknowledge the financial support provided by the Council for Scientific and Industrial Research (CSIR) and University Grants Commission (UGC), Government of India, respectively.

\end{document}